%% file: MeasuresInVisualizationSpace_compile.tex
\documentclass[graybox, envcountchap]{styles/svmult}

\usepackage{mathptmx}        
\usepackage{helvet}          
\usepackage{courier}         
\usepackage{outlines}         

\usepackage{makeidx}         
\usepackage{graphicx}        
\usepackage{multicol}        
\usepackage[bottom]{footmisc}

\input{preamble.tex}

\begin{document}

\input{properties.tex}
\boolfalse{createReferencesFromBBL}

\setcounter{chapter}{2}
\newcommand{\measurespreprint}{hhhhhhhhhhhhh}
\input{MeasuresInVisualizationSpace.tex}

\markboth{}{}
\author{}
\ifbool{createReferencesFromBBL}
{}
{\bibliography{MeasuresInVisualizationSpace}}
\bibliographystyle{styles/spmpsci}

\end{document}

%% file: properties.tex
\newcommand{\monthword}{\ifcase\month\or Jan\or Feb\or Mar\or Apr\or
                                         May\or Jun\or Jul\or Aug\or
                                         Sep\or Okt\or Nov\or Dec\fi}

%% file: MeasuresInVisualizationSpace.tex
\author{\\
\textsc{Fabian Bolte}\contact{\email{Fabian.Bolte@UiB.no}}, University of Bergen, Norway\\
\textsc{Stefan Bruckner}\contact{\email{Stefan.Bruckner@UiB.no}}, University of Bergen, Norway
~}

\chapter{Measures in Visualization Space}
\label{chp:Bruckner}

\abstract{Measurement is an integral part of modern science, providing the fundamental means for evaluation, comparison, and prediction. In the context of visualization, several different types of measures have been proposed, ranging from approaches that evaluate particular aspects of individual visualization techniques, their perceptual characteristics, and even economic factors. Furthermore, there are approaches that attempt to provide means for measuring general properties of the visualization process as a whole. Measures can be quantitative or qualitative, and one of the primary goals is to provide objective means for reasoning about visualizations and their effectiveness. As such, they play a central role in the development of scientific theories for visualization. In this chapter, we provide an overview of the current state of the art, survey and classify different types of visualization measures, characterize their strengths and drawbacks, and provide an outline of open challenges for future research.
}

\blfootnote{\\This is a preprint of a chapter for a planned book that was initiated by participants of the Dagstuhl Seminar 18041 (“Foundations of Data Visualization”) and that is expected to be published by Springer. The final book chapter will differ from this preprint.}

\par
\author{} 

\ifx\measurespreprint\undefined
    \newcommand{\measuresfigures}{parts/TheoryOfVisualization/chapters/MeasuresInVisualizationSpace/figures}
\else
    \newcommand{\measuresfigures}{figures}
\fi

\newpage
\section{Introduction}

Considering the vast amounts of data involved in many scientific disciplines, it is essential to provide effective and efficient means for forming a mental model of the underlying phenomena. Visualization seeks to provide these means through interactive computer-generated graphical representations, taking advantage of the extraordinary capability of the human brain to process visual information. Specifically, the term "visualization" refers to the process of extracting meaningful information from data and constructing a visual representation of this information. This process is composed of three basic stages~\cite{Janicke:2011:VSQ}

\begin{enumerate}
  \item making data displayable by a computer,
  \item transmitting visual representations to human viewers, and
  \item forming a mental picture about the data.
\end{enumerate}

Significant effort has been devoted to the formulation of taxonomies and categorizations of this general process. For instance, Shneiderman~\cite{Shneiderman:1996:EHI} introduced a task-by-data taxonomy, while Tory and M{\"o}ller~\cite{Tory:2004:RVH} focused on the classification of visualization algorithms. In an influential contribution, Munzner~\cite{Munzner:2009:NMV} proposed a nested model for designing and developing visualization pipelines, that has inspired a considerable amount of subsequent work. Wang et al.~\cite{Wang:2011:TFD}, for instance, proposed a two-stage framework for designing visual analytics systems, while Ren et al.~\cite{Ren:2013:MIM} proposed a multi-level interaction model of goal, behavior, and operation to facilitate system development with formal descriptions. The multi-level typology of Brehmer and Munzner~\cite{Brehmer:2013:MLT} distinguishes between the basic questions of \emph{why}, \emph{how}, and \emph{what}, in order to classify abstract visualization tasks. These types of classifications are highly valuable resources for visualization practitioners and researchers to steer the design process and to compare competing approaches.

\begin{figure}[tb]
  \centering
  \includegraphics[width=\textwidth]{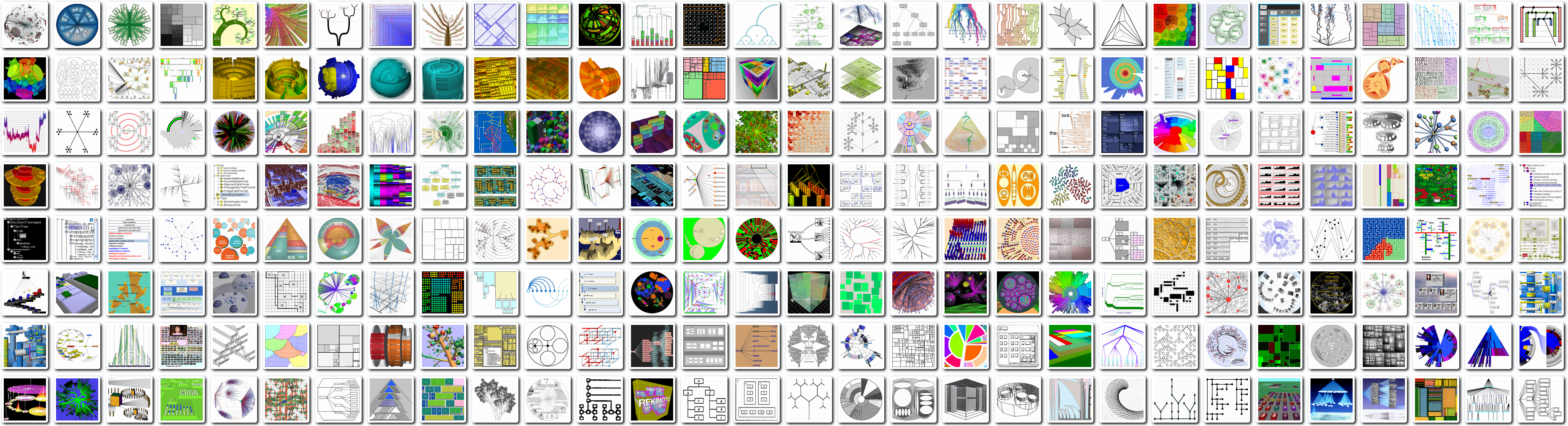}
  \caption{240 different tree visualization techniques~\cite{Schulz:2011:TTV} -- which one should be used?}\label{fig:treevis}
\end{figure}

Ultimately, however, in order to assess the effectiveness of visualization, it is crucial to know whether or not the mental picture of the data established by a human viewer is consistent with the original data, and whether or not one specific visualization technique or parameter setting is more effective than another. Displaying and analyzing data is of ever-increasing importance in almost all research disciplines. Consequently, the field of visualization is constantly growing and reliable visualizations are of more and more importance for domain experts to gain authentic insights. This progress comes along with a steady growth in diversity and complexity of visualization methods, making judgment of their effectiveness and suitability for a certain task difficult. Figure~\ref{fig:treevis} for instance, which shows 240 different techniques to visualize tree data taken from a visual bibliography on the topic~\cite{Schulz:2011:TTV}, illustrates the challenges in selecting appropriate visualization techniques.

Traditionally, visualization techniques and their parameter settings are evaluated by carrying out user studies which measure their performance for particular sets of tasks. However, such studies require considerable effort and their design is non-trivial~\cite{Munzner:2009:NMV}. Their specialized nature also makes it difficult to generalize the outcomes. Furthermore, when developing new visualization techniques, frequently only a small number of initial users is available, making it difficult to obtain statistically significant results. The alternative of solely relying on the visualization creator's judgment, is also scientifically questionable because it often reflects personal preference and may include bias. Hence, it is highly desirable to support a visualization process by enabling visualization creators to conduct an evaluation using objective measures.

In principle, such quality measures could then be used to automatically select and/or parameterize a visualization from a set of choices according to these measures by using an appropriate optimization process. Moreover, measures may also inform us about the structure of the visualization space itself, i.e., they may lead us to deeper insights into how the phenomenon of visualization works and hence could be of utility beyond a descriptive or evaluative usage. Hence, questions related to visualization measures are tightly connected to the bigger effort of specifying a theory of visualization. In this paper, we survey approaches that seek to enable the systematic analysis of visualization algorithms and their properties with respect to the underlying data characteristics and their perceptual qualities. While we cover the significant body of research that has been devoted to various types of visualization measures, we also specifically look at approaches that regard the interplay between data, algorithms and their parameters, and visual perception and cognition as a phenomenon that deserves study in its own right.

In many disciplines of science, hypotheses are formulated based on empirical data, and then subsequently developed into models and complete theories of the phenomenon under investigation. The predictions of these models and theories are then continuously validated and, once they are supported by sufficient data, are generally accepted as scientific "facts"\footnote{While a scientific theory can never be proven "true" in a mathematical sense, there are many examples of well-established theories such as evolution, quantum mechanics, general relativity, etc., that form the basis of modern science and that are rarely questioned on a principle level.}. Importantly, the consequences of these theories can lead to the discovery of new relationships and insights due to their predictions. Theoretical physics, for instance, heavily relies on the mathematical structure of existing well-validated theories in the development of more comprehensive models of our universe. There are many instances -- for example within the standard model of particle physics -- where subsequent discoveries have been predicted based on structural and mathematical aspects such as symmetries of the underlying theory. For instance, the famous Higgs mechanism and one of its important predictions, the Higgs boson, were already described in the 1960s, but strong evidence for its existence only became available in 2013.

The formulation of measures forms an important first step in the development of such theories, as they are often the fundamental building blocks from which more complex relationships can be derived. Thus, measures play a central role in the ongoing search for a more comprehensive theory of visualization.

\section{Measurement in Science}

In philosophy, the topic of measurement in science has been illuminated from many different points of view. Tal~\cite{Tal:2017:MS} gives a comprehensive account of the different schools of thought and here we will only briefly summarize his considerations in order to provide additional background. In principle, he distinguishes between the following perspectives:

\begin{enumerate}
  \item \textbf{Mathematical theories of measurement} regard measurement as the mapping of qualitative empirical relations to relations among numbers or other mathematical entities. Measurement theory aims to identify the assumptions related to the use of different mathematical structures for describing aspects of the empirical world. In particular, it attempts to make statements about the adequacy and limits related to the use of these structures. One of the key insights of measurement theory is that mathematical structures used for measurement should mirror relevant relations among the real-world objects being measured. For instance, we could mistakenly assume that an object measured at a temperature of 60 degrees Celsius is twice as hot as one measured at 30 degrees. However, when expressed using the Fahrenheit scale, the temperatures of these objects are 86 and 140, respectively. This is because the zero points of these two scales are arbitrary and do not correspond to the absence of temperature.

  \item \textbf{Realist views} consider measurement as the estimation of mind-independent properties and/or relations. A measurement is regarded as the empirical estimation of an objective property or relation. The term "objective", in this context, is meant to signify that these properties are independent of the conventions and beliefs of the humans conducting the measurement and of the methods used in their execution. The values of measurements are regarded as approximations of true values, and measurement itself is aimed at obtaining knowledge about properties and relations, rather than the assignment of values to objects themselves. For instance, a realist about length measurement would say that the ratio of the length of an object to the standard meter has a definite objective value, irrespective of how it is measured. The measurement itself is merely an approximation of this value.

  \item \textbf{Operationalist views} are concerned with the meaning and use of quantity terms. A realist would argue that these terms refer to sets of properties that exist independently of being measured. The operationalist point of view, on the other hand, is that the meaning of quantity concepts is solely determined by the set of operations used for their measurement. They view measurement as a set of operations that shape the meaning and/or regulate the use of a quantity-term. For example, length could be defined as the result of concatenating rigid rods, but it could also be defined by timing electromagnetic pulses. A strict operationalist would distinguish these two into distinct quantity concepts such as "length-1" and "length-2".

  \item \textbf{Information-theoretic accounts} view measurement as the gathering and interpretation of information about a system. Measuring instruments are regarded as "information machines" that interact with an object in a given state, encode that state into a signal, and convert this signal into an output. The accuracy of a measurement is dependent on the instrument as well as the level of noise in the environment. Information-theoretic accounts of measurement were originally developed by metrologists, and hence are practically oriented and tailored towards evaluating and improving the accuracy of measurement standards. As such, their connection to more philosophical considerations is less explored.

  \item \textbf{Model-based accounts} view measurement as the coherent assignment of values to parameters in a theoretical and/or statistical model of a process. According to model-based views, measurement consists of two levels: (1) a process involving interactions between an object of interest, an instrument, and the environment; and (2) a theoretical and/or statistical model (i.e., an abstract representations based on simplifying assumptions) that describes this process. Hence, the central goal of measurement is to assign values to the parameters of these models such that they satisfy certain criteria such as coherence and consistency.

\end{enumerate}

While these considerations are important and relevant lines of philosophical investigation, for the purposes of the discussion here we will largely gloss over these partially subtle distinctions. Nevertheless, we will see that some of these views are more prominent in the visualization domain than others. Many of the visualization quality measures are constructed in an operationalist manner, providing different means to measure the same property of a visualization. Several phenomena in visualization have been described by applying communication models from information theory, and several theoretical models try to explain, e.g., perceptual processes in the human visual system or the visualization process as a whole. Mathematical theories of measurement and realist views have received less attention in visualization research. As this topic gains more attention, we expect a more explicit exploration of the philosophical underpinnings of different approaches. In the following sections, we will describe different types of measurements in visualization and how they can be combined to build a better understanding of visualization as a research field in the future.

\section{Types of Visualization Measures}

There are numerous different aspects of the visualization process that one can set out to quantify. Partially, the boundaries between different types of measures can be fuzzy, but in the following we will attempt to characterize some principal categories of measures that have been investigated.

\subsection{Measures of Perceptual Characteristics}

The measurement of perceptual characteristics of visualizations aims to mimic low-level processing of visual stimuli in the human perceptual system. Essentially, the idea is that by -- at least partially -- modeling and simulating the early processing stages of the perception pipeline, we can predict how particular visual elements influence the interpretation of a particular visualization by a human observer.

Significant efforts have been devoted to understanding the effectiveness of different visual variables for encoding quantitative and qualitative data in the visualization literature. For example, Cleveland and McGill~\cite{Cleveland:1984:GPT} ran a well-known series of graphical perception experiments to measure accuracy in comparing values and to derive the rankings of encoding variables that still form the basis for many visualization design decisions. Similar types of experiments have also been used to compare different types of charts and their results have been employed to aid the automatic construction of visualizations ~\cite{Mackinlay:1986:ADG,Mackinlay:2007:SMA}.

A major early contribution to the study of visual perception was made by the Gestalt School of Psychology. Developed in the early 20th century, the intent was to understand the principles behind how humans acquire and maintain meaningful perceptions of the world given its complex and chaotic nature. The main idea maintains that the human perceptual system employs a notion of "gestalt" (German for shape or form) that it uses to organize and interpret its inputs. By further investigating this basic thought, psychologists were able to establish a series of Gestalt principles of perception, which are still respected today as accurate descriptions of visual behavior. Since then, several works have set out to describe these and related observations and their effects in a more formal manner.

At the most basic level, we can look at physiologically-based models which typically idealize neural behavior using mathematical functions. The response of retinal ganglion cells, which have a center-surround behavior, can be described by a difference-of-Gaussians function which contains a narrow excitatory center within a larger inhibitory surround~\cite{Rodieck:1965:QAC}. A Gabor function, mathematically defined as a 1D sinoid within a 2D Gaussian envelope, has been shown to be a good approximation of the edge patterns which the primary visual cortex (V1) neurons are sensitive to~\cite{Daugman:1985:URR}. Li~\cite{Li:1998:NMC} presented a model of contour perception in the primary visual cortex. While it does not include retinal processing or edge pattern recognition, it focuses on lateral connections in the visual cortex and how they can give rise to contour integration phenomena. Grossberg and Williamson~\cite{Williamson:2001:NMH} proposed a more detailed physiologically-based model which includes center-surround processing and Gabor-like pattern matching of neurons. It divides the primary visual cortex into several layers associated with particular behaviors such as contour enhancement and convergence of neural activity. Pineo and Ware~\cite{Pineo:2012:DVO} combine aspects of the models by Li and Grossberg and Williamson. They realize a difference-of-Gaussians retinal response and a V1 Gabor response. Furthermore, their approach is specifically tailored towards the viewing of data visualizations, which -- they argue -- tend to be viewed in an exploratory manner. Hence, they seek to model perception in the moments after viewing, before steady-state activity is reached. This also allows them to make the computational evaluation of the model sufficiently fast to be embedded in an optimization loop.
Thus, in addition to their model of low-level perception, Pineo and Ware~\cite{Pineo:2012:DVO} also present an application of their perceptual model for 2D flow visualization. They argue that the brain generates its high-level understanding of a visualization from the activity of low-level neurons, and erroneous low-level perception thus has a degrading effect on this high-level understanding. Based on this reasonable assumption, they propose a predictor for the perceived direction at a point in visual space from the activity of edge selective neurons that surround it. Likewise, they predict the perceived speed of flow from the activity of blue-yellow neurons (which correspond to their chosen color mapping) weighted by the distance of the receptive field to the point being predicted. These measures are then used in a hill climbing optimization process to adjust the parameters of a streaklet-based visualization.

Such perceptual measures focus on the low-level processing of visual stimuli in the human perceptual system such as preattentive processing~\cite{Healey:2012:AVM}. Hence, they are primarily concerned with how basic visual encoding variables, such as position, length, area, shape, and color, and the interaction of the variables (e.g., integrable or separable), influence the efficiency of low-level perceptual features such as visual search, change detection, and magnitude estimation~\cite{Behrisch:2018:QMI}. While physiological models taking into account neural response are scientifically attractive due to their "first-principles" nature, an obvious challenge is to scale them up to more informative aspects of higher-level perception. As is the case in many areas of science, it is far from trivial to connect multiple scales in a meaningful manner while preserving important practical aspects such as computational feasibility. For this reason, the modeling of higher-level phenomena often ignores some of the more detailed aspects. In the context of perceptual measures, the concept of saliency~\cite{Itti:1998:MSV} is a prominent example for this.

In general, visual saliency models assess the features of an image to predict which areas of that image will draw a viewer's attention. While they are typically inspired by the structure and function of the human visual cortex and are designed to be "biologically plausible", most approaches make a number of simplifying assumptions. Several practical saliency models have been proposed that, while inspired by basic principles such as the center-surround mechanism, forego more detailed modeling of the neural response and instead take a more phenomenological approach. Saliency models can be categorized as models of bottom-up visual attention. Bottom-up visual attention is drawn to regions that are distinct from their surroundings with respect to their basic visual features such as contrast, color, or motion. Top-down visual attention, on the other hand, is driven by the viewer's goals, expectations, and experience. It is hence allocated voluntarily based on the viewer's task and prior knowledge~\cite{Connor:2004:VAB,Pinto:2013:BTA}. This makes saliency an attractive basic task-agnostic measure for investigating how viewers read a visualization in principle and thus saliency-based measures have garnered the interest of visualization researchers.

Kim  and  Varshney~\cite{Kim:2006:SEV}, for instance, presented a method that enhances the saliency of selected regions in volumetric data which they validated using an eye-tracking study. Lee et al.~\cite{Lee:2005:MS} applied the concept of saliency to surface meshes and showed how the measure can be used for targeted simplification as well as viewpoint selection. J{\"a}nicke and Chen~\cite{Jaenicke:2010:SQM} proposed an approach which uses a saliency-based metric to measure the mismatch between data-space feature maps and the visual representation of the data. While most types of saliency models are tailored towards natural scenes, Matzen et al.~\cite{Matzen:2018:DVS} developed a method specifically targeted at abstract data visualizations.

Overall, perceptual measures are a useful tool for determining and/or predicting which parts of a visualization will be most prominently seen by a user. Combined with an appropriate way to characterize relevant features in the data, they can be utilized to detect potential mismatches between the importance of regions in data space and their perceptual prominence in the final image. However, at present only low-level perceptual processing can be feasibly taken into account and higher-level aspects or even cognition are still beyond the reach of current approaches.

\subsection{Task-Oriented Quality Measures}

In contrast to lower-level perceptual measures, the goal of quality measures is to inform about the performance of a visualization technique with respect to a particular well-defined task assumed to be important for the overall goal of the visualization. As discussed in the survey by Behrisch et al.~\cite{Behrisch:2018:QMI}, a particular characteristic of such measures is that they do not explicitly consider the user. Instead, they often attempt to heuristically quantify the presence and/or extent of an "anti-pattern", i.e., an assumed known defect or undesirable characteristic of a visualization. These types of measures are commonly referred to as "quality metrics" in the visualization literature. However, as pointed out by Behrisch et al.~\cite{Behrisch:2018:QMI}, this is a somewhat misleading term as "metric" has a precise meaning in mathematics with well-defined properties (i.e., non-negativity, identity of indiscernibles, symmetry, and the triangle inequality) which need not necessarily hold in all cases. Thus, we adopt the more neutral term "measure" which does not have these implications.

\begin{figure}[tb]
	\centering
	\includegraphics[width=\textwidth]{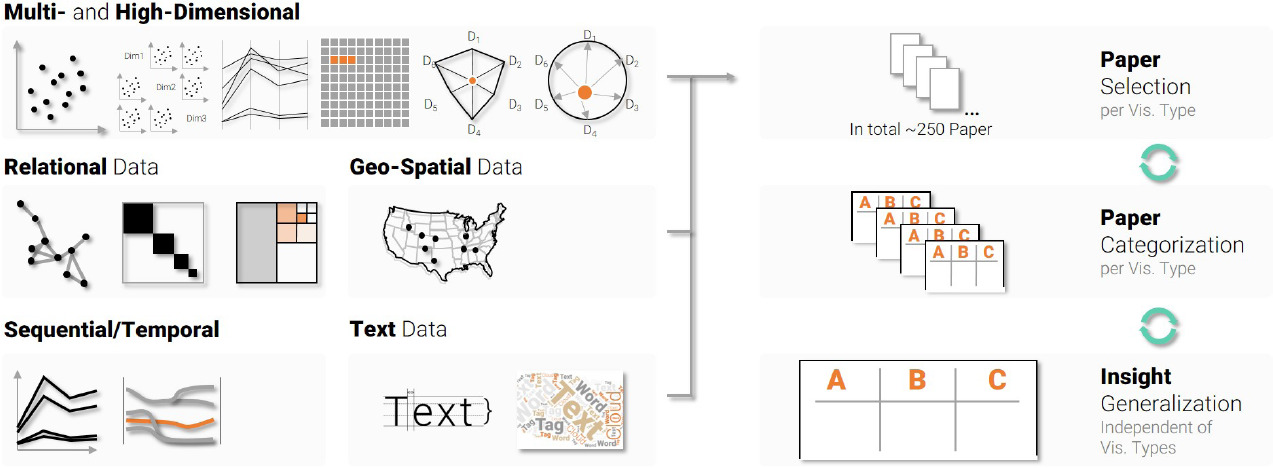}
	\caption{Behrisch et al.~\cite{Behrisch:2018:QMI} analyzed and categorized quality measures from around 250 papers in visualization. These mid-level measures are mostly specific to the underlying data, task, and visualization technique.}
	\label{fig:qualitymetrics}
\end{figure}

As the recent state-of-the-art report by Behrisch et al.~\cite{Behrisch:2018:QMI} focuses on these types of measures (classified as "mid-level perceptual quality metrics" in their work), we will only briefly summarize well-known approaches and refer the reader to their comprehensive survey for further details. Given their specialized nature, it makes sense to discuss task-oriented quality measures according to the type of visualization they are designed for, as shown in Figure~\ref{fig:qualitymetrics}. For instance, in scatterplots and scatterplot matrices "scagnostics" -- based on an idea by Tukey and Tukey~\cite{Tukey:1985:CGE} -- have been introduced as an approach to identify anomalies based on attributes of their shape and appearance. These measures themselves form a multi-dimensional space which can be explored in a scatterplot matrix in order to identify outliers in the form of unusual scatterplots. Wilkinson et al.~\cite{Wilkinson:2005:GTS} later presented graph-theoretic methods to implement the same approach using a set of measure categories (outliers, shape, trend, density, and coherence), each composed of multiple numerical measures. For example, the shape of scattered points in a plot can be described by the following measures: convexity, skinniness, stringiness, and straightness. Bertini and Santucci~\cite{Bertini:2004:QMS,Bertini:2005:ISE} proposed a model for visual clutter in scatterplots based on an estimate of colliding points vs. available space. They subsequently derived a quality measure that aims to quantify whether the relative data density is preserved when considering the represented density in the plot. It is also common for quality measures to be defined implicitly, for example as part of a layout algorithm. For instance, Byron and Wattenberg~\cite{Byron:2008:SGG} presented an approach to optimize the appearance of stacked graphs by using measures such as deviation and wiggle.

Task-oriented quality measures have arguably received the most attention in the field of visualization, as they often tend to encode -- with varying degree of fidelity -- known best practices or common shortcomings specific to a particular class of visualizations. In essence, they can be seen as (partial) formalizations of design recommendations, and thus tend to be quite practically oriented. Typically being grounded in well-established principles in visualization makes these types of measures semantically meaningful and expressive, providing a sound basis for optimization as well as for the comparison of different but related algorithms. A potential downside of this applied nature of task-oriented quality measures is their limited generalizability.

For instance, when rendering streamlines in flow visualizations, there exist different seeding strategies to define the starting points and number of streamlines. The overall goal is to display all features in the flow without introducing clutter. This goal introduces a trade-off between increasing the number of streamlines to cover all features, but decreasing it for better clarity. While this optimization is crucial for streamline visualizations, it is so specific that it can hardly be applied to any other type of visualization. This is true for many task-specific quality metrics. As a consequence, whenever a new visualization technique is discovered, new task-oriented quality measures need to be developed to optimize the specific aspects of this visualization. It would be desirable to define general measures that express the quality of a visualization independent of its type and allow for their comparison.

As an example, edge crossings can be optimized for multiple visualizations, i.e., graphs, parallel coordinates, and storylines. This is because all of these techniques utilize edges (or links) as a visual encoding for aspects of the underlying data. The user's ability to read a chart is influenced by the number of edge crossings as well as the angle at which they cross, and there seem to be higher-level perceptual aspects of such visual embeddings that increase the cognitive load on the user. If we manage to define these aspects instead of task-specific features, then we might be able to form a more general theory about the perception of visualizations. This would not only allow us to compare different techniques on an equal basis, but further enable the prediction of how new visualization techniques will perform given their defined visual mapping.

\subsection{Structure-Oriented Measures}

In contrast to task-oriented measures, this class attempts to quantify general structural elements of the visualization process. More specifically, structure-oriented measures aim to express in a -- at least in principle -- measurable form fundamental characteristics of the visualization process itself.
Classical examples for these types of measures are Tufte's data-to-ink ratio, as well as his lie factor~\cite{Tufte:2001:VDQ}. The former describes the proportion between the amount of pixels used to present data and the total amount of pixels, whereas the latter describes the ratio between the size of a data value and the size of its corresponding visual element. Both express desirable relationships between the data and its visual representation, but are not tied to any particular visual encoding. On the contrary, they aim to describe general qualities of visualizations, and thus play a particularly important role in considerations towards a theory of visualization.

Mackinlay~\cite{Mackinlay:1986:ADG} was one of the first to discusss the expressiveness and effectiveness of visualizations as general means to compare and choose different visual designs. He describes expressiveness as the ability to encode all facts of a dataset without introducing additional facts that are not in the data. Effectiveness, on the other hand, further depends on the user's capabilities to read a certain visualization. Having the user introduced as a deciding factor for the effectiveness of a visualization requires a detailed understanding of the human visual system and, although a lot of research has been contributed towards this goal, we still do not possess a sufficiently complete model that would allow us to predict this on a general level. We are therefore further reliant on empirical results of user studies to describe the perceptual capabilities of visualization users.

Demiralp et al.~\cite{Demiralp:2014:VEM} evaluated visual mappings in general by assessing how well the input data is represented by visual elements. They describe visualization as a function that maps from a domain of data points to a range of visual primitives. They further argue that the same measures that can be found in data, like symmetry and distance, should be reflected in the visual elements. In this sense, we could encode pairwise difference in data space as pairwise perceptual difference in color, shape, size, or others. One problem with this approach is that perceptual distance is not given in most visual spaces and needs to be estimated empirically. Additionally, we often utilize several visual encodings at the same time, and it is unclear how they interact and potentially interfere. The authors argue that when two visual spaces are combined, a measure for that space can be constructed from the individual measures. When acquiring perceptual measures for all kinds of visual spaces, we could then create a standard library to validate the pairwise distances between elements in all kinds of visualizations.

Inspired by these considerations, Kindlmann and Scheidegger~\cite{Kindlmann:2014:APV} argue that distance functions and metrics have limits, since for example partial orders are not symmetric. They instead developed an algebraic framework for describing symmetries between manipulations in data space and their resulting consequences in visualization space. From this, they derived three principles that should be true for a mapping from data to visualization, i.e., unambiguous data depiction, representation invariance, and visual-data correspondence. In short, the visual mapping should make sure that a change in the data is reflected by a corresponding change in the visualization, while changes in the data representation (e.g., the specific data structures used in the implementation) do not affect the visualization, and significant changes in data should result in noticeable changes in the visualization. Given some examples, it becomes clear that not always all of these principles can be met. The visualization designer needs to be aware of certain shortcomings and make sure that the right principle is respected given the task at hand. In this work, the authors introduced a uniform description of different design choices. They adhered to a mathematical model that describes the process of visualization based on its structural properties. They further mention that user studies can be utilized to test perceptual distinguishability and thereby complement mathematical models. The conjunction of evaluated visualizations and  mathematical models can help to make statements about visualizations which are not yet evaluated through user studies. While this approach still relies on some notion of perceptual distance, it is notable as it does lead to measurable predictions that in principle can be verified without reliance on user studies. For instance, it can be tested without user involvement whether a significant change in the data leads to no change in the visualization. This opens up the door for a set of "unit tests" for visualization, which could verify at least some objective characteristics fully automatically.

Information theory has been a major influence in the search for solid theoretical foundations in visualization. Silver~\cite{Silver:1995:OOV} employed the concept of object orientation to conceptualize the visualization process, arguing that the definition and abstraction of features into objects, and their interactions in local regions, allows for a better measurability of phenomena and understanding of their evolution. Abstracting the features of a scientific domain into such a concept allows for generally applicable measurements such as volume, diameter, and curvature and provides a basis for objective comparison. Jankun-Kelly et al.~\cite{Kelly:2007:MFV} proposed the P-Set model to describe a user's interactions as choosing a parameter based on a previous parameter set, and applying the new set to derive a transformed visualization result. As demonstrated by Liu et al.~\cite{Liu:2008:DCT}, distributed cognition can be utilized as a theoretical framework in visualization. Purchase et al.~\cite{Purchase:2008:TFI} analyzed which existing theoretical models can be applied to visualization and provided suggestions for their integration. In particular they considered visualization under the light of data-centric predictive theory, information theory, and scientific modeling. Chen and J{\"a}nicke~\cite{Chen:2010:ITF} applied information theory to describe phenomena in visualization with communication models. They argued that many problems and features in visualization can be explained by similar phenomena from information theory which can be applied to evaluate visualizations on a more general level. Xu et al.~\cite{Xu:2010:IFF} followed a similar idea to evaluate visualizations by measuring the amount of information that is transported through the visual channels and applied this framework to flow visualization examples. Wang and Shen~\cite{Wang:2011:ITS} complemented this work by additional principles with a particular focus on scientific visualization. Category theory and semiotics were employed by Vickers et al.~\cite{Vickers:2012:UVF} to facilitate an improved understanding of visualizations in practice and to describe a well-formed visualization process. The conceptual framework of visual multiplexing by Chen et al.~\cite{Chen:2014:VM} facilitates the study of different mechanisms for integrating and overlaying multiple pieces of visual information.

Based on these information-theoretic considerations, Chen and Golan~\cite{Chen:2015:WMV} introduced a comprehensive cost-benefit model of visualization, defining cost as the search space for answers. They utilized the big O notation to classify tasks accordingly. Presenting a fact or piece of information has cost $O(1)$, observations as in "What happened?" require the user to read all data points, which has a complexity of $O(n)$. When looking into correlations, causes, and other complex relationships, we must consider a broader spectrum of relations, ending up at $O(n^k)$. And, finally, when we want to derive a model for visualization, taking into account all parameters and algorithmic steps, the complexity might be $O(n!)$.
They further introduced a cost function, which can be derived from energy, time, or monetary measurements necessary to find the answer. They defined benefit as a gain in certainty about the information. Based on these definitions, they derived an incremental cost-benefit ratio that describes the amount of effort required to compress the information towards the point that the user's initial question can be answered and a decision can be made. Based on this formulation, it is in principle possible to use an optimization process to discover the best visualization method.

Bruckner et al.~\cite{Bruckner:2018:MSD} proposed a model to analyze the directness of interaction techniques in visualization. They considered the different mappings involved in the visualization process, i.e., the mapping from data space via the visualization space to the output space (e.g., a monitor or a head mounted display), as well as the subsequent perceptual and cognitive processes involved in generating the user's mental model. They then investigated the parallel process of interaction, starting from an intended action (based on the user's mental model) via the manipulation space (i.e., a physical interaction device such as a computer mouse) to the interaction space and finally back to the data space. Based on this model, they introduced a measure for the degree of indirectness of an interactive visualization setup based on how invertible the involved mappings are and demonstrated how this measure can be practically realized.

Compared to task-specific quality metrics, describing visualizations on a general level not only provides us with a better understanding of visualization as a scientific research field, but further allows us to make predictions about non-evaluated, or even not yet developed visualization techniques. For instance, when the interaction with the visualization does not coincide with gathered knowledge about interaction directness, the user is likely to experience a discrepancy between their intended and executed manipulation. One major question is how the sheer number of theoretical frameworks and models can be combined and integrated into one coherent knowledge base. Similar to other research fields like physics, where theories about electricity and magnetism have been combined into a larger theory of electromagnetism, visualization could gather greater insights by combining existing theoretical frameworks, leading to a fundamental strengthening of the research field as a whole.

\subsection{Meta-Perceptual Process Measures}

So far we have primarily examined well-established and generally accepted measures and models to evaluate visualization with the goal of optimizing task execution time, easing data exploration, or increasing the gained insight. But when visualization is utilized as a knowledge source for the general public, we can formulate other equally important goals for visualization design. In education, we might be interested in creating memorable knowledge or engage students in working with a visualization. In commercial scenarios, aspects such as aesthetics and impact, or even the profitability of a visualization can be the main goals of a specific design. We summarize these higher-level aspects as meta-perceptual process measures that aim to characterize additional qualities that go beyond what are typically considered to be primary desired properties of a visualization in the research community. In some sense, such measures aim to capture the attributes of a visualization from the point of view of other domains, such as art or economics.

\begin{figure}[tb]
	\centering
	\includegraphics[width=1.\textwidth]{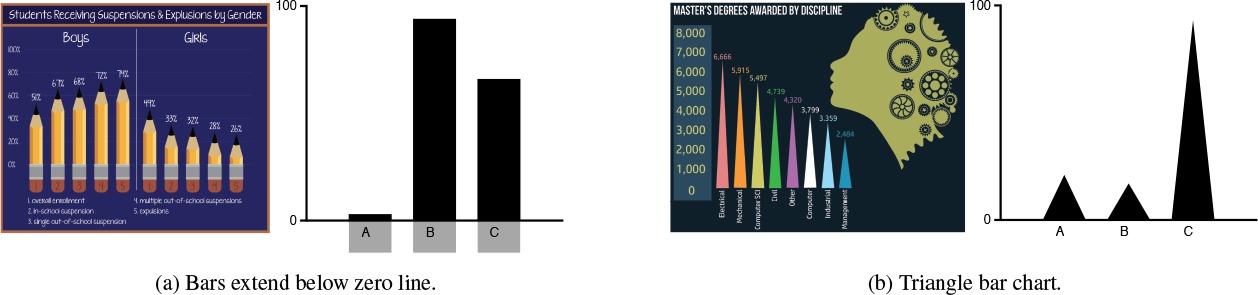}
	\caption{Comparison of plain and embellished bar charts~\cite{Skau:2015:EIV}. Even small visual modifications, like using triangular instead of rectangular bar charts, will increase the error rate. Embellished representations can increase the memorability of the visualization.}
	\label{fig:embellishment}
\end{figure}

For instance, Healey et al.~\cite{Healey:1996:HSV} conducted experiments to evaluate how hue and orientation allow users to accurately estimate features in visualizations through preattentive processing. The question was if a short glimpse at a visualization can convey the general message, and if it can, which factors influence this capability. While Skau et al.~\cite{Skau:2015:EIV} showed that even small visual embellishments increase the error rate when reading bar charts, Bateman et al.~\cite{Bateman:2010:UJE} found that visually embellished visualizations are more memorable than plain charts. Figure~\ref{fig:embellishment} shows two visual mappings for bar charts, as well as their embellished counterparts. Borkin et al.~\cite{borkin2013makes} investigated which elements of visualizations make them memorable. They showed, for example, that color, human recognizable objects, high visual density, and unique design improve the ability of humans to remember a visualization. Furthermore, memorability was independent of subjects' context and biases.

Aesthetics of a visualization are hard to measure and in most cases subjective. Tractinsky et al.~\cite{Tractinsky:2000:WBU} found a strong correlation between aesthetics and usability, which suggests it as an important factor for designing and evaluating visualizations. Lau and Moere~\cite{Lau:2007:TMI} proposed a model for aesthetics in information visualization, seeing aesthetics as the degree of artistic influence on the data mapping, rather than as a measure of appeal. Filonik et al~\cite{Filonik:2009:MAI} summarized several possible measures of aesthetics for information visualization from the literature and concluded that many aspects of this phenomenon remain unexplored. Harrison et al.~\cite{Harrison:2015:IAD} ran a user study and found correlations between certain measurable visual features and visually appealing aesthetics. They found that colorfulness and visual complexity have a positive correlation to perceived aesthetics, but depend on gender, age, and level of education.

Saket et al.~\cite{Saket:2016:BUP} summarized and reviewed several of these meta-perceptual criteria in the field of visualization. They described engagement as the amount of time spent with the visualization, proposed a model for measuring enjoyment~\cite{Saket:2015:TUE}, and found that pictorial representations and embellished visualizations increase enjoyment~\cite{Saket:2016:CNL}. Their work concluded that memorability, engagement, and enjoyment are complex aspects of visualizations that are hard to quantify, and require further study. It is, for example, not yet clear how interactions affect these measures, and many more factors that influence a user's experience might exist.

A somewhat different class of measures is related to non-cognitive aspects of visualization. For instance, van Wijk~\cite{Wijk:2005:VV} proposed a model to measure the "profitability" of a visualization in an economic sense. In this model, the cost of a visualization (e.g., development cost and users' time to understand the visualization) is considered in relation to the return on investment in the form of knowledge gain. The value of a visualization can thus be increased if many people use it regularly, obtain valuable knowledge, and spend less time or money to make a decision. Unfortunately, knowledge gain is a rather broad and vague concept, so more precise notions are needed to quantify this aspect more accurately.

Compared to previously discussed approaches, meta-perceptual process measures have so far mostly been evaluated in rather narrow scenarios, providing guidelines for visualization design. More quantitative measures that would allow for the comparison of different visualizations with respect to the outlined qualities have not been explored extensively. Furthermore, the fact that some qualities like aesthetics are not necessarily directly related to common visualization goals such as the generation of insight, may have lead some visualization researchers to discard them as irrelevant. However, we believe that it is important to also consider the impact of visualization in a broader context, and hence find that the measurement of such properties is an important and worthwhile endeavor. Parallels may be drawn to other fields -- for instance, organizational performance was once mostly viewed in terms of its economic characteristics, but organizational psychology has shown that measures of occupational health and well-being such as job satisfaction can be important predictors for the financial success of a company.

\section{Towards a "Bigger Picture"}

\begin{figure}[tb]
	\centering
	\includegraphics[width=.80\textwidth]{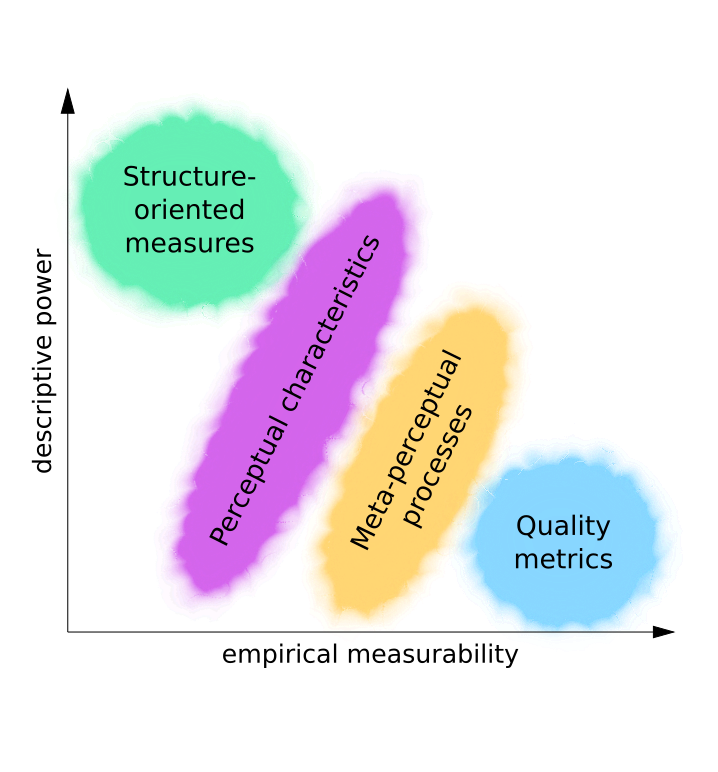}
	\vspace*{-14mm}
	\caption{Overview of visualization measures regarding their ease of being measured and their capability of describing visualizations as a whole. While quality metrics are easy to measure, they are in most cases too specific to find applicability in generalized observations. Meta-perceptual measurements try to capture more general, higher-level phenomena, but require user studies to be quantified. Although they are applicable to a large range of visualization techniques, their generated insight follows specific intents (like making a visualization memorable). Perceptual studies try to understand the human visual system and could, if fully understood, explain many phenomena in the analysis of visualizations. At present, however, where only the low-level visual processing is well understood, their applicability is limited to rankings of visual channels and encodings for rather isolated situations. Theories about visualization are among the most general and descriptive approaches for describing visualizations. Although some of them propose varying measures for the quality of a visualization and allow for their comparison, they are in many cases still too abstract to be applied in practical use cases.}
	\label{fig:overview}
\end{figure}

As can be seen from previous examples, there is still a long way to go towards quantitative statements about visualizations in general. Figure \ref{fig:overview} provides a high-level overview of the discussed measure categories with respect to their practicality as well as their ability to describe general phenomena. Many of the presented quality measures are specific to a certain type of visualization, like wiggle in streamgraphs, or scagnostics for scatterplots. Counting the number of edge crossings in a visualization is an example that can be applied to several different visualization techniques, like graphs, streamgraphs, and parallel coordinate plots, but is still specific to visualizations that utilize visual links for their layout. It could be argued, that a meter can measure width, height, and length in the real world, because every object has to have these properties, given their underlying molecular structure. Visualizations, on the other hand, utilize a number of visual properties to encode varying information, even encoding semantically similar information with different visual encodings. From this point of view, it is no surprise that different subareas in visualization have developed vastly varying quality measures. Kosara~\cite{Kosara:2016:EBS} looks at many of the best practices followed in visualization and encourages researchers to build a better, well justified basis for knowledge about visualizations.

Some properties, like clutter, empty space, and overplotting are more general and can be used to characterize visualizations on a more fundamental level. But their effect on the users' perception varies and is therefore often evaluated through user studies. Several examples, as for instance discussed by Harrison et al.~\cite{Harrison:2015:IAD}, have shown that measures which by themselves do not make a statement about quality (e.g., colorfulness and visual complexity) can be transformed into quality measures, when evaluated with a user study. The users' perception can be measured or quantified and thereby operate as an indicator for quality. The fact that this type of evaluation can be performed for visualizations in general means that there might be a common ground that allows for comparability. While Behrisch et al.~\cite{Behrisch:2018:QMI} provided an excellent summary of task-dependent quality measures, a survey of existing studies would be able to provide an overview of what these studies have in common and on how specific they are to their individual task and visualization types.

In order to compare not only different visual encodings, but visualization types, we would require a standardized way of evaluating common properties. For instance, different visualizations might apply the same color map and when asking users of different visualizations the same questions we would acquire comparable answers. One major problem of this approach is that visualizations are, among other things, data-, task-, and user-dependent. While a given dataset might create clutter in one visualization, it might not in another, and the opposite can be true for yet another dataset. Some visualizations are better in giving an overview, while others provide detailed insights, and the questions asked in user studies are often task-dependent to investigate exactly these specific strengths or weaknesses. When asking task-independent questions in order to keep the results comparable, insight on these specific differences might get lost. Lastly, visualizations can be targeted towards a certain audience, being more specific for experts, or more intuitive for the broader public. For this reason, participants of a user study are often chosen from the specific audience, introducing a bias towards the background and knowledge the participants have. If the goal is to create comparable results, the distribution of participants would need to be as general as possible, introducing additional problems like participants not having the background knowledge required to, e.g., benefit from a visualization in the medical domain.
Independent of the mentioned shortcomings, we might be able to come up with some general statements that provide insight into the users' mental model and opinion about the visualization given their data and purpose they operate on, similar to the System Usability Score introduced by Brooke~\cite{Brooke:1996:SUS}. The interpretation of such a score, as in this case demonstrated by Bangor et al.~\cite{Bangor:2008:EES}, can lead to a description of the general performance of a visualization, and bring us closer to a common basis for comparability across fields.

Perceptual studies in particular provide more general means by analyzing the human vision and ranking different visual channels based on their capability of presenting information. They build a fundamental understanding of the basic principles of visualization and are applicable to all kinds of visualization types. So far, we merely understand low-level perceptual processes. This fact limits the applicability of perceptual studies to make general statements about visualizations and predict their usability. Meta-perceptual metrics, on the other hand, try to evaluate higher-level features independent of the specific visual encoding. Aesthetics, engagement, and enjoyment have a major impact on the way users interact with the visualization and on how the gained knowledge is memorized. Despite several efforts taken in this direction, these measures have mostly been explored in information visualization and require further research in other fields like, e.g., scientific visualization. When we have a better understanding of how these phenomena behave in different visualization types, we can build a more general theory and learn from the insights gained.
In addition to already mentioned measures, proxy measures can be used to quantify properties that are otherwise hard to observe. The idea is to find a measurable property that strongly correlates with the phenomenon we want to analyze.

    As a reflection of a discussion panel on how to pursue theoretical research in visualization, Chen et al.~\cite{Chen:2017:PTA} described different evaluation approaches and how they can contribute to a theoretical foundation. Taxonomies classify objects of interest, such as data types, visual encodings, user tasks, or interaction techniques into groups and subgroups. Ontologies then describe additional relationships between these different groups and entities, providing a more detailed picture of the underlying interactions. Guidelines describe the quality of a certain approach and make statements about which practices should or should not be used in order to achieve a desired outcome. The authors argue that guidelines need to be evaluated and refined over time, as well as transformed into quantitative laws when applicable. {VisGuides}~\cite{Diehl:2018:VisGuides, VisGuides} provides a platform to openly discuss guidelines in visualization and allow for their continuous refinement. When a guideline has shown to be useful over the years, it can be established as a principle. Conceptual models describe a general idea or understanding of how certain processes or systems work in order to reason about their structure and functioning. For example, a perceptual model describes how we think the human visual system works, which allows us to derive conclusions and best practices, although we have not fully understood this system yet. Such models can further be supported by mathematical frameworks, like information theory. In our opinion, the combination of quantitative measures and a mathematical framework can form the basis of more general models of visualization. These can then be used to reason about causal relationships and make testable predictions. We believe, that the main goal of our community should be to unify existing approaches into larger theories about visualization that incorporate acquired knowledge into a more general understanding of the subject itself. Sacha et al.~\cite{Sacha:2014:KGM} demonstrated how perceptual and theoretical frameworks, as well as guidelines, can be combined into a model for understanding the process of knowledge generation. We should continue this line of thought to further integrate quality and meta-perceptual measures into theoretical frameworks and to create general models of the visualization process. By continuously verifying and refining these models, we can continuously advance visualization theory and strengthen the research field for greater accomplishments to come.

\section{Conclusion}

Other research fields have shown how incremental refinement and verification of theoretical models can lead to major leaps in knowledge and understanding. In visualization, we have seen several promising attempts towards a theoretical foundation, as well as greater acknowledgement and presence of theoretical papers. We can learn from other scientific disciplines and bear in mind that the formulation of a theory and definition of measures in visualization do not need to be perfect from the very beginning. Practical barriers, like not being able to compute a measure due to technical limitations, should not prevent us from suggesting and formulating such concepts. Many important milestones in scientific history, like Einstein's general relativity or Feynman's quantum electrodynamics, have been postulated much earlier than they could be verified. Similarly, Fermat's Last Theorem took 358 years from its proposition to a mathematical proof. Such theories allow us to state our assumptions, formulate predictions, and develop technological advances, even if they are not well-verified or "proven" yet. Evaluation efforts can be made not only to assess specific visualization techniques or applications, but to empirically test theories. Based on such continuously validated and refined theories, we are optimistic that we will eventually be able to evaluate and compare visualization techniques on a more general level, predict how users will perceive and interact with the visualization, and develop new visualization techniques for better decision making.

\section*{Acknowledgments}
The authors would like to thank all participants of the Dagstuhl Seminar for constructive and fruitful discussions. This work was supported by the MetaVis project (\#250133) funded by the Research Council of Norway.